# Probing Population III Initial Mass Functions with He II/Hα Intensity Mapping

Jasmine Parsons[1,2], Lluís Mas-Ribas[1,3], Guochao Sun[3], Tzu-Ching Chang[1,3], Michael O. Gonzalez[3], and Richard H. Mebane[4,5]
[1] Jet Propulsion Laboratory, California Institute of Technology, 4800 Oak Grove Drive, Pasadena, CA 91109, USA; jasmine.parsons@mail.mcgill.ca
[2] McGill University, Department of Physics & McGill Space Institute, 3600 Rue University, Montréal, QC H3A 2T8, Canada
[3] California Institute of Technology, 1200 E. California Blvd., Pasadena, CA 91125, USA
[4] Department of Astronomy and Astrophysics, University of California, Santa Cruz, 1156 High Street, Santa Cruz, CA 95064, USA
[5] Department of Physics and Astronomy, University of California, Los Angeles, CA 90024, USA

## Abstract

We demonstrate the potential of line-intensity mapping to place constraints on the initial mass function (IMF) of Population III stars via measurements of the mean He II 1640 Å/Hα line-intensity ratio. We extend the `21cmFAST` code with modern high-redshift galaxy-formation and photoionization models, and estimate the line emission from Population II and Population III galaxies at redshifts $5 \leqslant z \leqslant 20$. In our models, mean ratio values of He II/Hα $\gtrsim 0.1$ indicate top-heavy Population III IMFs with stars of several hundred solar masses, reached at $z \gtrsim 10$ when Population III stars dominate star formation. A next-generation space mission with capabilities moderately superior to those of CDIM will be able to probe this scenario by measuring the He II and Hα fluctuation power spectrum signals and their cross-correlation at high significance up to $z \sim 20$. Moreover, regardless of the IMF, a ratio value of He II/Hα $\lesssim 0.01$ indicates low Population III star formation and, therefore, it signals the end of the period dominated by this stellar population. However, a detection of the corresponding He II power spectrum may be only possible for top-heavy Population III IMFs or through cross-correlation with the stronger Hα signal. Finally, ratio values of $0.01 \lesssim$ He II/Hα $\lesssim 0.1$ are complex to interpret because they can be driven by several competing effects. We discuss how various measurements at different redshifts and the combination of the line-intensity ratio with other probes can assist in constraining the Population III IMF in this case.

*Unified Astronomy Thesaurus concepts:* Initial mass function (796); Population III stars (1285); Stellar populations (1622); Line intensities (2084); Cosmology (343)

## 1. Introduction

Population III stars are theorized to be the first stars, formed in the early universe from metal-free gas, and to have played a key role in driving the beginning of the epoch of reionization (Barkana & Loeb 2001; Bromm 2013). Despite their crucial nature, however, many of their basic properties, such as their initial mass function (IMF), are still unknown. Initial studies suggested that Population III stars may present top-heavy IMFs with many more massive stars than one would expect in more typical IMFs (e.g., bottom-heavy, Salpeter (Salpeter 1955) IMFs), with stellar masses up to several hundred or a thousand solar masses (e.g., Abel et al. 2002; Bromm et al. 2002). More recent simulation work, however, found that their masses could be as low as a few tens of solar masses or less (e.g., Hosokawa et al. 2011; Stacy et al. 2012, 2016; Hirano et al. 2014, 2015; Susa et al. 2014). Low mass results are also in line with inferences from Galactic archeology considering elemental abundance patterns of extremely metal-poor stars (e.g., Ishigaki et al. 2018). This large range of potential masses results in tremendous differences when considering the impact of Population III stars on the process of reionization and cosmic metal enrichment, yet the Population III IMF remains unconstrained.

Even with upcoming instruments such as the James Webb Space Telescope (JWST; Gardner et al. 2006), direct observations of Population III galaxies at redshifts above $z \sim 10$ are challenging (e.g., Zackrisson et al. 2011b; Mas-Ribas et al. 2016; Schauer et al. 2020). Detections by means of gravitational lensing (Zackrisson et al. 2012, 2015; Windhorst et al. 2018; Vikaeus et al. 2022) or the combination of several observational techniques such as wide-field surveys and multiband photometry (Inoue 2011; Zackrisson et al. 2011a) appear to be possible, but in these cases the number of detected sources may not be representative of the entire stellar population.

In order to gain information about the overall Population III stellar population at high redshifts, Visbal et al. (2015) proposed the use of the line-intensity mapping (LIM) technique (Madau et al. 1997; Suginohara et al. 1999; Visbal & Loeb 2010). Rather than resolving individual sources, line-intensity mapping statistically measures the aggregate luminosity of a particular line over a large patch of the sky. Owing to its nature, therefore, this technique is not limited to bright sources above a given luminosity threshold (see the review by Kovetz et al. 2017). Intensity mapping has been proposed and is being conducted for a variety of emission-line frequencies such as [C II] 158 $\mu$m (e.g., Gong et al. 2012; Silva et al. 2015; Pullen et al. 2018; Yang et al. 2019; Yue & Ferrara 2019; Chung et al. 2020; Sun et al. 2021b), CO (e.g., Righi et al. 2008; Gong et al. 2011; Lidz et al. 2011; Keating et al. 2016, 2020; Li et al. 2016; Chung et al. 2019; Ihle et al. 2019; Breysse et al. 2022), 21 cm (e.g., Scott & Rees 1990; Chang et al. 2008, 2010; Switzer et al. 2013; Anderson et al. 2018), and several other bright lines driven by star formation, such as hydrogen Lyα, Hα, Hβ, [O II] 3727 Å, or [O III] 5007 Å (e.g., Silva et al. 2013; Pullen et al. 2014; Comaschi & Ferrara 2016; Fonseca et al. 2017; Gong et al. 2017, 2020; Heneka et al. 2017; Croft et al. 2018; Mas-Ribas &

Chang 2020). However, Visbal et al. (2015) suggested the less-used He II 1640 Å line, which is particularly interesting for studies of Population III galaxies owing to its sensitivity to the hardness of the ionizing spectrum incident on the emitting gas. The energy required to ionize the He II ion is 54.4 eV, four times higher than that required to ionize H I. Normal-type (Population II) galaxies typically emit little He II 1640 Å due to the low average energy of their ionizing photons. On the other hand, the lack of metals and the high masses and temperatures of Population III stars can result in high energy ionizing photons able to produce large amounts of He II 1640 Å emission (Schaerer 2002, 2003; Raiter et al. 2010; Mas-Ribas et al. 2016).

In this paper, we extend the work of Visbal et al. (2015) by narrowing in on the Population III IMF. Specifically, we consider the He II 1640 Å/H$\alpha$ line-intensity ratio as a tracer of the Population III IMF, first proposed by Oh et al. (2001), and investigate its detectability with intensity mapping over a range of scenarios and redshifts. Unlike He II, the amount of nebular H$\alpha$ emission is not particularly sensitive to the hardness of the ionizing stellar spectrum (Schaerer 2002, 2003). Therefore, H$\alpha$ plays the role of a normalization factor in the line-intensity ratio that accounts for different amounts of stellar mass and/or star formation when comparing different IMFs.

For our calculations, we use and extend the semi-numerical code 21cmFAST (Mesinger & Furlanetto 2007; Mesinger et al. 2011). 21cmFAST is suitable for intensity mapping calculations because it allows us to perform large-scale computations exploring different reionization parameterizations and scenarios in a short timescale. Using a semi-numerical approach is beneficial because it allows us to include detailed (nonlinear) physical effects and dependencies between model parameters beyond simpler analytical prescriptions.

This paper is organized as follows: in Section 2, we describe our simulations and IMFs. The mean line-emission results for He II and H$\alpha$, as well as their corresponding ratio, are presented in Section 3. The detectability of these two individual power spectrum signals and their cross-correlation are subsequently explored in Section 4. Finally, we discuss the interpretation of the He II and H$\alpha$ signals and the challenges of line interloper separation in Section 5, before concluding in Section 6.

Throughout this work, we assume a flat, $\Lambda$CDM cosmology consistent with Planck 2016 (Ade et al. 2016).

## 2. Simulations

For the present study, we use a precursor version of our recently developed simulation tool LIMFAST (L. Mas-Ribas et al. 2022, in preparation; G. Sun et al. 2022, in preparation). Here, we consider a simulation volume of 300 comoving Mpc with cell size of 1.5 cMpc on a side, and we adopt the galaxy models of Furlanetto et al. (2017), which describe feedback-regulated star formation in galaxies growing through continuous accretion onto dark matter halos. Unlike the Furlanetto (2021) models implemented in LIMFAST, the Furlanetto et al. (2017) approach does not consider the detailed treatment of interstellar gas. However, our choice is motivated by the fact that in the current work we examine nebular lines whose emission can be connected to the star formation of the host galaxy without depending directly on the specific amount and properties of the galactic gas. The Furlanetto et al. (2017) prescriptions, therefore, should suffice for our purposes, and we do not explore further improvements or comparisons between galaxy models. Below, we focus on the characteristics of the code used here and refer interested readers to the above references for more details on LIMFAST.

Our simulation builds upon and extends the semi-numerical code 21cmFAST (Mesinger & Furlanetto 2007; Mesinger et al. 2011), and takes advantage of some of the implementations by Park et al. (2019). Our extensions include the aforementioned high-redshift galaxy-formation models (Section 2.1) by Furlanetto et al. (2017), and the calculation of emission-line radiation related to star formation (Sections 2.2 and 2.4) via precomputed photoionization calculations with Cloudy (version 17.02, Ferland et al. 2017). Furthermore, we introduce a redshift-dependent treatment for the production of ionizing photons used in the ionization calculations during reionization (Section 2.3). We briefly describe the main characteristics of 21cmFAST next and then elaborate on our modifications.

The 21cmFAST code begins by creating evolving density and velocity fields from a set of initial conditions via the application of first-order perturbation theory (Zel'Dovich 1970). Next, the code computes the amount of collapsed matter at each cell by means of the conditional Press-Schechter formalism (Press & Schechter 1974; Lacey & Cole 1993). For this calculation, 21cmFAST normalizes the Press-Schechter halo-mass function in such formalism to match the Sheth-Tormen halo-mass function (Sheth & Tormen 1999) including the correction by Jenkins et al. (2001). We stress that this approach corresponds to the "matter density field" case in 21cmFAST and not the "halo finder" case. This implies that individual halos are not identified or resolved in the simulation. Instead, we adopt and make use of the halo-mass function, whose amplitude at each cell is driven by the amount of collapsed matter and, in turn, the overdensity value in that region. Similarly, the code does not populate the halos by creating galaxies. The galaxy properties, such as star formation, luminosity, etc., depend on the halos through the star-formation model described in the next section, and their average cell values are obtained by integrating them over the halo-mass function corresponding to each cell. The ionization state of the baryonic gas is then computed by comparing the cumulative number of ionizing photons from radiation sources and the number of neutral hydrogen atoms in spherical regions, from large to small volume sizes (analogously to the excursion set formalism Furlanetto et al. 2004), which allows for the computation of inhomogeneous reionization scenarios. The 21 cm brightness temperature is finally obtained from the spin temperature by using the previously calculated density, ionization, and thermal state of the gas, and the hydrogen Ly$\alpha$ radiation background fields (see Mesinger & Furlanetto 2007; Mesinger et al. 2011, for details).

### 2.1. Star Formation

Our simulation starts by modeling the process of star formation in galaxies, departing from the 21cmFAST procedure, once the halo distribution has been set. We parameterize the star formation via the star-formation rate, $\dot{M}_\star(M, z)$, which we assume to depend on the host halo mass and redshift. Other quantities of interest are then derived from this star-formation rate.[6]

---

[6] This simple treatment allows users to simulate intensity mapping signals by implementing any custom galaxy model that gives a relation between star formation and halo mass and redshift.

Following Furlanetto et al. (2017), we use the accretion rates of individual halos of a given mass and redshift, $\dot{M}(M, z)$, computed from abundance matching by the code ARES (Mirocha et al. 2017), and derive the star-formation rate as

$$\dot{M}_\star(M, z) = f_\star(M, z) \frac{\Omega_b}{\Omega_m} \dot{M} . \quad (1)$$

Here, $\Omega_b/\Omega_m \approx 1/6$ is the ratio of baryons to dark matter, and $f_\star$ is the fraction of baryonic gas that turns into stars, i.e., the star-formation efficiency, which we parameterize following the momentum-driven feedback scenario described by Furlanetto et al. (2017). Following the prescriptions by these authors,

$$f_\star = \frac{f_{\text{shock}}}{f_{\star,\text{max}}^{-1} + \eta(M, z)}, \quad (2)$$

where $f_{\star,\text{max}} = 0.2$ is the maximum star-formation efficiency in the momentum-driven feedback model. The numerator in the previous equation,

$$f_{\text{shock}} \approx 0.47 \left(\frac{1+z}{4}\right)^{0.38} \left(\frac{M}{10^{12} M_\odot}\right)^{-0.25}, \quad (3)$$

represents the fraction of gas that is prevented from being accreted onto galaxies due to the virial shock, having been adopted by Furlanetto et al. (2017) from the hydrodynamical simulations of Faucher-Giguère (2011) for the baryonic mass assembly of halos, with an upper limit of $f_{\text{shock}} = 1$. The term

$$\eta(M, z) = C \left(\frac{10^{11.5}}{M}\right)^\xi \left(\frac{9}{1+z}\right)^\sigma, \quad (4)$$

denotes the relation between the rate of gas expelled from the galaxy and the star-formation rate, where we adopt the parameter values $C = 5$, $\xi = 1/3$, and $\sigma = 1/2$ that give rise to an $f_\star$ value consistent with the results from Sun & Furlanetto (2016) based on halo abundance matching.

We note that there exist star-formation prescriptions in the literature (e.g., Silva et al. 2015; Behroozi et al. 2019) other than the one we are using here. We expect, however, our choice of star-formation model to not impact the He II/H$\alpha$ ratio to first order, since this choice would affect the two individual line signals in the same manner. Furthermore, we will later describe the simple expression that we use to model the transition from Population III to Population II stars. This expression does not depend on any specific star-formation prescription, and thus the line-intensity ratio would be unaffected by changes in the star-formation model. If this Population III to Population II transition were instead connected to specific properties of the galaxies, however, then different star-formation models may impact the resulting line-intensity ratio.

Once the relation between star-formation rate and halo mass is obtained, and with the halo-mass function (HMF) calculated by 21cmFAST in a simulated region, one can calculate the star-formation rate density (SFRD) as

$$\dot{\rho}_\star(z) = \int \dot{M}_\star(M, z) dn/dM(M, z) dM . \quad (5)$$

The minimum halo mass in the above integral is assumed to be the atomic cooling halo mass computed by 21cmFAST, which ranges from $\sim 10^7 M_\odot$ at $z = 20$ to $\sim 10^8 M_\odot$ at $z = 5$. The integral upper limit extends to infinity but in practice the code stops the calculation at $M_{\text{max}} = 10^{16} M_\odot$. The number density of halos at high redshift above this threshold is small and, therefore, the exact value does not change the main results. The term $dn/dM(M, z)$ represents the default HMF used in 21cmFAST, consistent with the Sheth-Tormen (Sheth & Tormen 1999) HMF with the correction by Jenkins et al. (2001).

Finally, using Equation (23) in Furlanetto et al. (2017), we define the average metallicity in a simulated volume as

$$Z(z) = \frac{y_Z \rho_\star}{\bar{\rho}_b f_{\text{coll}}(z)}, \quad (6)$$

where $\rho_\star$ denotes the accumulated stellar mass density obtained by integrating $\dot{\rho}_\star(z)$ over all redshifts to account for the buildup of metals over time. The term $y_Z = 0.01$ above denotes the fraction of stellar mass that is returned to the gaseous medium in the form of metals, $\bar{\rho}_b$ is the mean cosmic baryon density, and $f_{\text{coll}}$ implies that the metals reside in collapsed structure (i.e., galaxies). We note that this metallicity expression contains the integrated stellar mass and not the integral over the HMF as in the previous equations because it also includes the terms denoting the mean baryon density and the mean collapsed fraction in the entire simulation box. The use of these terms implies that the metallicity obtained in this way is valid when accounting for a large population of objects or a large volume (a simulation cell) but not for individual halos. For the same reason, the luminosities derived from this metallicity should be considered at a cell level and not for individual halos as is the case in LIMFAST. For our current purpose, this simplified approach is valid because we are most interested in Population III galaxies that do not depend on metallicity (we assume they are metal free), and because we consider average values over the cell when estimating the signal of Population II galaxies. Finally, for simplicity, we assume the stellar and gas metallicity to be the same.[7]

### 2.1.1. Population III Star Formation

To characterize the Population III sources, we use a simple functional form that sets their star formation to a fraction, $f_{\text{Pop III}}$, of the total cosmic star-formation rate at a given redshift. Specifically, the fraction of total cosmic star formation that is Population III is given by

$$f_{\text{Pop III}}(z) = \frac{1}{2}\left[1 + \text{erf}\left(\frac{z - z_t}{\sigma_p}\right)\right], \quad (7)$$

with the fiducial transition redshift $z_t = 14$ and transition width $\sigma_p = 4$. For comparison, in Section 3, we will also assess the impact of an alternative choice of parameter values with $z_t = 18$ and $\sigma_p = 2$, which yields a significantly reduced amount of Population III star formation. While the exact star-formation history of Population III stars is highly uncertain given the lack of observations, these choices of parameters are broadly consistent with previous works taking into account indirect observational constraints and theoretical arguments of feedback regulation as shown by Sun et al. (2021a). The simple prescription adopted here implicitly assumes that Population III stars form in the same halo-mass range as Population II stars, instead of in small molecular cooling halos alone. Population III star formation in massive halos is still a viable scenario considering that the mixing of metals released by Population III supernovae is inefficient, as well as the potential effects of the radiation field on the regulation of star formation (e.g., Xu et al. 2013, 2016; Visbal et al. 2014, 2016). Because we are most interested in the impact of the IMFs on the range of line-intensity ratio values and less on their detailed redshift dependence, we do not account for more sophisticated star-formation prescriptions in this work (see, e.g., Wise et al. 2014; Mebane et al. 2018; Visbal et al. 2018; Schauer et al. 2019; Qin et al. 2020, 2021; Muñoz et al. 2022; Tanaka & Hasegawa 2021), for some recent physically motivated Population III star-formation recipes, as well as implementations into `21cmFAST`. Figure 1 displays the evolution of $f_{\text{Pop III}}$ with redshift for the fiducial and alternative cases described above. As illustrated in this figure, the main difference between the two scenarios is an earlier and faster decrease of $f_{\text{Pop III}}$ with time for the alternative case.

---

[7] Possible small differences between stellar and gas-phase metallicity can exist in practice in our code. These differences may arise because of the discrete sampling of metallicity values used in our stellar spectra, while the gas-phase metallicity can take any value within a continuous range. In any case, these differences are $\Delta \log_{10}(Z) \leqslant 0.05$, which denotes the half distance between metallicity steps in the discrete case. Because this value is smaller than the potential deviations induced by non-solar $\alpha$/Fe abundances (Steidel et al. 2016) we do not perform further corrections.

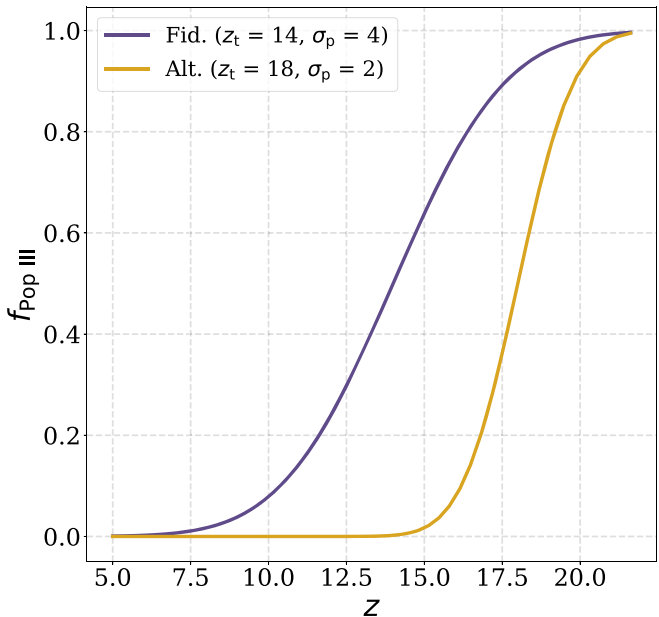

**Figure 1.** Redshift evolution of the fraction of star formation in the form of Population III stars considered in this work (Equation (7)). The fiducial functional form shown in purple has a transition redshift of $z_t = 14$ and a width of $\sigma_p = 4$. The alternative function shown in golden transitions at $z_t = 18$ with a width of $\sigma_p = 2$.

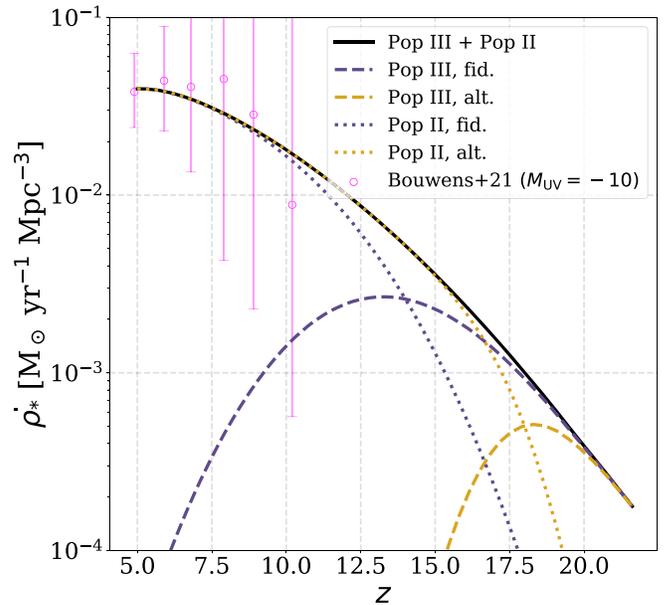

**Figure 2.** Evolution of the star-formation rate density with redshift for Population II (dotted lines) and Population III (dashed lines), and for the fiducial (purple lines) and alternative (golden lines) $f_{\text{Pop III}}$ evolutions. The overall star-formation rate density with redshift is denoted by the solid black line. Driven by the evolution of $f_{\text{Pop III}}$, Population III stars in the alternative case dominate only at the highest redshifts and for a short time compared to the fiducial scenario. The magenta data points with error bars denote the observational data by Bouwens et al. (2021), extrapolating the luminosity functions down to $M_{\text{UV}} = -10$ for comparison.

With the prescriptions for the cosmic star formation presented in Section 2.1 and the Population III fraction described above, one can now obtain the star-formation histories of Population II and Population III galaxies. Figure 2 shows the evolution of the SFRD with redshift for Population II (dotted lines) and Population III (dashed lines) stars, in the fiducial (purple lines) and alternative (golden lines) $f_{\text{Pop III}}$ models. The redshift evolution of the total SFRD is denoted by the solid black line and the error bars represent the observations by Bouwens et al. (2021) for comparison, extrapolating the UV luminosity functions down to $M_{\text{UV}} = -10$. This brightness threshold broadly corresponds to the minimum halo masses considered in our calculations. The error bars illustrated here represent the maximum and minimum values allowed by the combination of parameter uncertainties in the Bouwens et al. (2021) luminosity functions, and they are not the $1\sigma$ uncertainties. Driven by the $f_{\text{Pop III}}$ evolution, Population III galaxies in the alternative case dominate the star formation for a short time at very high redshift, while for the fiducial case their impact is notable down to lower redshifts. This trend will have direct implications for the measurement of the He II/H$\alpha$ line-intensity ratio described in later sections.

### 2.2. Stellar and Nebular SED Modeling

The original `21cmFAST` code considers one spectral energy distribution (SED) to describe all the Population II galaxies and another one for Population III stars, and uses them for the calculation of the radiation fields. Below, we describe our newly implemented SEDs for Population III stars (Section 2.2.1), and a set of metallicity-dependent SEDs created for Population II stars (Section 2.2.2).

#### 2.2.1. Population III SEDs

For the Population III stellar SEDs, we use the hydrogen and helium-only stars at the zero-age main sequence (ZAMS) phase computed by Mas-Ribas et al. (2016).[8] We refer the interested reader to that work for details on the calculation of the spectra.

---
[8] The SEDs are available at https://github.com/lluism/seds.

**Table 1**
Population III IMFs Used in This Work

| IMF | $m_{\min}$[a] | $m_{\max}$[a] | $\alpha$ | $Q$ (H I)[b] |
|---|---|---|---|---|
| m9M100a0 | 9 | 100 | 0 | 6.5 |
| m9M100a2 | 9 | 100 | 2.35 | 2.6 |
| m9M500a0 | 9 | 500 | 0 | 13.5 |
| m9M500a2 | 9 | 500 | 2.35 | 5.0 |
| m120M500a0 | 120 | 500 | 0 | 13.7 |
| m120M500a2 | 120 | 500 | 2.35 | 12.0 |

**Notes.**
[a] The lower and upper mass limits are in units of $M_\odot$.
[b] The photon flux are in units of $10^{47}\,\mathrm{s}^{-1}\,M_\odot^{-1}$.

Following the methodology in Mas-Ribas et al. (2016), we create here six stellar populations covering different stellar mass ranges and IMF slopes, which will be used to explore their impact on the line-emission signals. These stellar populations, which we refer to as IMFs for simplicity, are described in Table 1. The second and third columns in the table denote the lower and upper mass limits of the IMFs, respectively, and the parameter $\alpha$ in the fourth column is the slope in our adopted power-law IMF expression equating

$$\xi(m)\,dm \propto m^{-\alpha}\,dm, \tag{8}$$

where $\xi(m)$ gives the differential number of stars with mass in the range $m \pm dm/2$. The values $\alpha = 0$ and $\alpha = 2.35$ represent top-heavy and Salpeter (Salpeter 1955) IMFs, respectively. Finally, the fifth column denotes the rate of hydrogen-ionizing photons produced per unit of solar mass. To summarize our IMF writing convention in Table 1, "m" refers to the lower mass bound, "M" refers to the upper mass bound, "a" refers to the power-law slope, and the stellar mass is in units of $M_\odot$, following the style in Mas-Ribas et al. (2016).

Following the method in `21cmFAST`, we tabulate the information of the SEDs for each IMF by computing the number of photons per stellar baryon within the first 23 energy levels of the hydrogen atom (Barkana & Loeb 2005). Furthermore, for each particular IMF case, we use for the ionization computations the number of ionizing photons per stellar baryon of Population III stars that corresponds to the Population III SED describing such IMF for self-consistency. We assume that the SEDs describe the Population III stellar populations during 3 Myr in all cases, consistent with the lifetime of Population III stars with masses $\sim 120\,M_\odot$. The production rate of ionizing photons of our SEDs differs by a factor of two to six compared to those used by Barkana & Loeb (2005), who considered a Population III SED from Bromm et al. (2001) with stars in the mass range $100 \leqslant m/M_\odot \leqslant 1000$, and with a stellar lifetime of about 2 Myr. The exact value of this difference depends on the IMF considered. The lifetime value assumed here is somewhat arbitrary but it is degenerate with a potential duty cycle for the Population III galaxies, and with the (also highly unknown) escape fraction of photons. In other words, this lifetime value of 3 Myr is not equivalent to stating that the stars live 3 Myr. Rather, it represents an "effective" time that encompasses various unconstrained effects such as a potential Population III duty cycle and the escape fraction of photons.

The luminosities of the emission lines used here are computed by using our Population III SEDs as the incident spectrum in the photoionization code `Cloudy` (version 17.02, Ferland et al. 2017), together with a set of parameters describing the gaseous (nebular) medium. In particular, we consider a spherical geometry with a distance of $r_{\mathrm{in}} = 10^{19}$ cm between the nebular gas and the central source, a gas density of $n_{\mathrm{H}} = 100\,\mathrm{cm}^{-3}$, and a metallicity $Z/Z_\odot = 10^{-6}$, where $Z_\odot = 0.014$ denotes the solar metallicity (Asplund et al. 2009). We adopt this low metallicity value for all redshifts and cases when considering Population III stars to describe an almost metal-free environment. Finally, we set the ionization parameter in `Cloudy` to a fiducial value of $U = 10^{-2}$, but we explore other values in Appendix A.

We will perform simulations considering one IMF for each realization, corresponding to one Population III SED, and we will explore the impact of their properties on the mean He II/H$\alpha$ line-intensity ratio in Section 3. Finally, the line intensities and number of photons yielded by the Population III components will contribute to the total budgets by the fractional amount corresponding to the value of $f_{\mathrm{Pop\,III}}$ at each redshift.

### 2.2.2. Population II SEDs

The implementation of Population II SEDs is similar to that of Population III, but here we create a set of 41 Population II SEDs computed with `CloudyFSPS` (Byler et al. 2017, 2018). Each SED represents a stellar (and gas) metallicity value between $\log(Z/Z_\odot) = 1$ and $\log(Z/Z_\odot) = -3$, with a step size of $\Delta \log_{10}(Z) = 0.1$. Cells with metallicities outside this range are assumed to have the corresponding limiting value. In practice, however, `CloudyFSPS` quotes a metallicity range of $-2 \leqslant \log(Z/Z_\odot) \leqslant 0.2$, and we find that the number of photons computed beyond these limits is indeed almost unchanged. This limitation implies a maximum underestimation (when the cell metallicity is $\log(Z/Z_\odot) = -3$) of $\sim 40\%$ in the number of He II photons, and less for H$\alpha$ (Schaerer 2003). Owing to this small effect, and because Population II stars play a secondary role in our work, we do not pursue further corrections to these SEDs. We highlight, however, that care must be taken when using `CloudyFSPS` for reionization studies where galaxy metallicities can be notably low. The age of the stellar populations is set to 3 Myr considering bursty star formation, and we assume an average stellar lifetime of also 3 Myr for the production of ionizing photons. These time values are again degenerate with the escape fraction of photons and/or galaxy duty cycles but they do not have a major impact on our results for line emission. Furthermore, we use a typical Salpeter IMF with a mass range of $0.1$–$100\,M_\odot$ to describe the Population II stellar population.

For the computation of the total emission-line luminosity, we consider the contribution from the luminosity in each cell. For a specific cell, the luminosity is obtained by linearly interpolating the luminosity values of the two SEDs with metallicity values the closest to that of the corresponding cell. The interpolation is performed over the tabulated emission-line results for each Population II SED, where the corresponding SEDs were used as the incident spectrum in `Cloudy`, with the same nebular parameters detailed above and where the we considered the metallicity of the gas and the stellar SED to be the same.

### 2.3. Ionizing Emission

A further implementation in our code is an improved treatment of the ionization calculations. The `21cmFAST` code uses two different numbers of ionizing photons per baryon, one for Population II and one for Population III stars, regardless of redshift. Our code, instead, computes a number of ionizing photons per baryon that depends on the SEDs, and that evolves with redshift driven by the metallicity evolution. In detail, we use the number of ionizing Population III photons that represents the IMF of interest, and the number of Population II photons that corresponds to the Population II SED with a metallicity value closest to the mean metallicity value in the simulation box at a given redshift. In practice, the number of ionizing photons with contributions from both stellar populations is obtained as $N_{ion,Z} = N_{ion}^{III} f_{Pop\,III} + N_{ion}^{II}(Z)(1 - f_{Pop\,III})$, where the redshift dependence is carried by the metallicity $Z(z)$. Using the same number of ionizing photons for all sub-volumes in the simulation box at one time step is still a simplified approach, but we adopt it as is from `21cmFAST` for computational simplicity (see Davies & Furlanetto 2022, and references therein for improved treatments of the ionization calculations). Finally, although the line emission depends weakly on our assumptions for the ionization calculations because we consider an ionizing escape fraction of ⩽0.1, we discuss the ionization histories resulting from our models in Appendix B.

### 2.4. Line Emission

Considering the aforementioned ingredients, now we can detail the calculation of line emission. We define the line-luminosity density in a given cell and redshift as

$$\rho_l(z) = t_e\, \dot{\rho}_\star(z)\, l(Z, U), \qquad (9)$$

where $l(Z, U) = l^{III}(U) f_{Pop\,III} + l^{II}(Z, U)(1 - f_{Pop\,III})$ is the luminosity per stellar mass obtained from interpolating the tabulated values, and accounting for the contribution from Population II and Population III stars. The term $t_e = 3$ Myr denotes the assumed duration during which the line emission occurs in galaxies. We have set this same value here and in the radiation computations described just above for simplicity. In reality, however, $t_e$ can be different for the ionization and the line-emission calculations, for the Population III and Population II cases, and even for the specific emission line of interest. For line emission, we have tested that $t_e = 3$ Myr produces similar Population II luminosities as the same SEDs computed assuming a continuous star-formation mode over a period of a few tens of megayears. We do not perform more detailed assessments of this value here and note again that this parameter is degenerate with the duty cycle of galaxies, the escape fraction of line emission, which we here assume to be 100%, and that our line-emission results scale linearly with $t_e$.

Finally, the observed specific intensity is derived from the luminosity density as

$$I_\nu(z) = \frac{c}{4\pi} \frac{\rho_l(z)}{\nu_0\, H(z)}, \qquad (10)$$

where $\nu_0$ is the rest-frame frequency of the emission line of interest, $c$ is the speed of light, and $H(z)$ is the Hubble parameter at redshift $z$.

The mean line intensity at a given redshift is the mean intensity of the cells in the simulation box at that redshift, which has a side length of 300 comoving Mpc. We can then divide the mean line intensities of He II and H$\alpha$ to obtain the He II/H$\alpha$ ratio values presented in the next section.

## 3. He II and H$\alpha$ Emission

Below, we present the results of our fiducial simulations, and discuss variations on the fiducial model in Section 3.1.

The top and middle panels in Figure 3 show the evolution with redshift of the mean H$\alpha$ and He II line intensities, respectively, for our six IMFs described in Table 1. The mean intensity of the H$\alpha$ line generally increases toward low redshift following the evolution of the cosmic star-formation rate (Figure 2) for all IMFs. However, the gradual disappearance of Population III stars produces a change in the evolution, most visible as a peak at $z \sim 12$ for the case of the most top-heavy IMFs. In this scenario, there is a decrease in total intensity because the difference in the H$\alpha$ production from Population III and Population II stars is large (Schaerer 2003) and the effect of reducing the Population III fraction is not readily compensated by the increase of Population II star formation. For the faintest Population III IMFs, the transition between the epochs dominated by Population III and Population II emission is less visible. This behavior occurs because the faint Population III IMFs have emissivities more similar to those of Population II than the hardest Population III IMFs, and therefore the change is smoother. A local maximum around the transition redshift can also be observed in the mean He II intensity evolution. He II, however, is mostly produced by Population III stars and is thus more affected by the decrease in $f_{Pop\,III}$ than H$\alpha$. The six curves of He II also do not yet converge at low redshift, unlike those of H$\alpha$, due to this sensitivity.

The bottom panel of Figure 3 shows the mean ratio of the H$\alpha$ and He II intensities. In this panel, the m120M500a0 IMF curve (solid orange) and the m9M500a0 IMF curve (solid gray) are nearly indistinguishable at a ratio value of He II/H$\alpha \approx 0.1$, with a maximum difference of ∼4% between them. This is because the ratio is dominated by the presence of massive stars and, therefore, it is most sensitive to the upper mass bound and slope of the IMF. For the same reason, the Salpeter IMFs (dashed curves) always have lower ratio values than their harder-spectra top-heavy counterparts (solid curves). In this case, the difference is most extreme for the IMFs with the largest range of masses (i.e., the largest difference between the upper and lower mass limits) because the overall mass in our calculations is the same for each IMF calculation and, therefore, a larger range implies fewer of the most massive stars.

The same panel also shows that the fainter the Population III IMF, the higher the redshift required to measure an informative ratio concerning the IMF of Population III stars without the contaminant effect of Population II emission. As mentioned in Section 1, the potential tracer for the Population III IMF is the He II/H$\alpha$ ratio from *only* Population III stars. Emission of He II and H$\alpha$ from Population II stars is thus a contaminant to the ratio we are investigating. This contaminating effect can be seen in the ratio panel of Figure 3. When Population III stars dominate at high redshifts, the ratio is flat, however the decline of the Population III and rise of the Population II stellar populations cause the ratio to decline at different slopes depending on the IMF. The effect of the Population III IMF at these lower redshifts would thus be degenerate with the Population II SFR in our calculations, making a measurement of the He II/H$\alpha$ ratio no longer a simple indicator of the Population III IMF. In detail, the ratio of the top-heavy Population III IMFs can be reliably measured at $z \gtrsim 12$, while, for the faintest IMFs, a measurement should be performed at $z \gtrsim 15$.

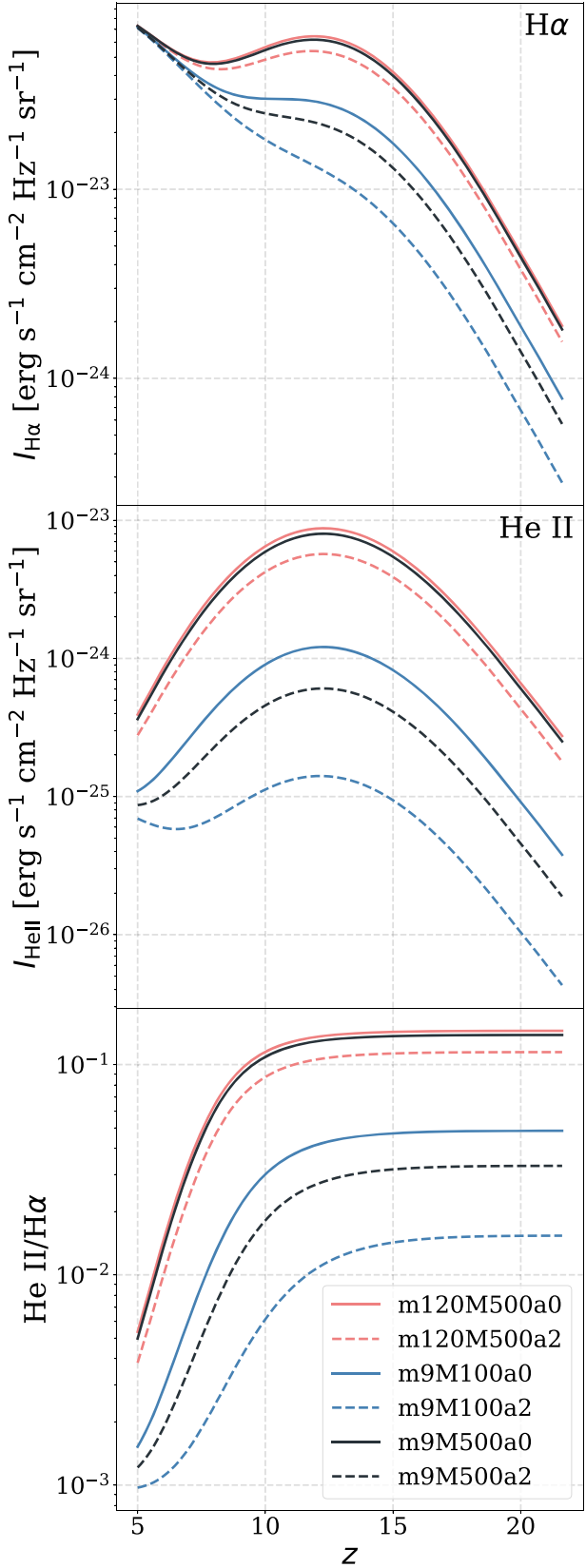

**Figure 3.** Mean line-intensity evolutions with redshift for H$\alpha$ (top) and He II (middle), and their ratio (bottom), using the fiducial simulation parameter values described in Section 2 and the six Population III IMFs described in Table 1. The disappearance of Population III stars and the impact of the Population II stellar population is most notable at $z \lesssim 12$.

In conclusion, a measurement of the mean line-intensity ratio presenting values of He II/H$\alpha \gtrsim 0.1$ would suggest top-heavy IMFs with upper mass limits in the hundreds of $M_\odot$. Lower ratio values complicate the interpretation because they may be affected by several causes. A lower value can indicate lower upper mass limits, different IMF slopes, or measurements affected by a strong Population II component. Furthermore, the exact value of the ratio will change when there exist differences between the photon escape fraction of the two lines.

### 3.1. Variations on the Fiducial Model

Figure 4 illustrates the same calculations as Figure 3, now with a fraction of Population III star formation described by an alternative $f_{\text{Pop III}}(z)$, with center at $z_t = 18$ and a width of $\sigma_p = 2$. While H$\alpha$ and He II still present peaks in their intensity evolutions, these are shifted to higher redshifts and contracted along the redshift axis when compared to the fiducial case. This is expected given the change in the position of the center and the width of the underlying error function. Furthermore, their intensity levels are also shifted down around an order of magnitude, consistent with the lower overall Population III population in these alternative simulations. The convergence of the six Population III IMF curves into a single curve is also more evident here at $z \sim 13$, since Population III stars disappear more rapidly in this alternative scenario; after no Population III stars remain, the line intensities increase in time as a power-law curve following the cosmic star-formation history. Due to the early disappearance of Population III stars, the ratio of He II to H$\alpha$ also decays earlier and quicker than its fiducial counterpart. The need to go to high redshift for a clean, Population III-dominated measurement of this ratio is once again emphasized. It is worth noting, however, that measuring the value of the ratio at low redshift also yields interesting information. A ratio below $\sim 10^{-2}$ indicates that Population III stars have almost disappeared regardless of the IMF (contributing to the total star formation by only a few percent). Therefore, investigating the redshift where this level is reached lends an insight onto the transition from Population III to Population II stars.

In addition to changing the underlying error function describing the evolution of $f_{\text{Pop III}}$, we investigate the impact of variations in the values of the ionization parameters used in our calculations on our line-intensity evolutions. Since these effects are relatively small, they are discussed in Appendix A.

In summary, the mean He II/H$\alpha$ ratio is indeed sensitive to the underlying Population III IMF, and particularly to the upper mass bound of the IMF. Reliable ratio measurements have to be performed at redshifts high enough for Population III stars to dominate the signal and avoid the contamination by Population II stars that may reduce the true ratio value. Measurements of mean ratio values of $\lesssim 10^{-2}$ are also informative because they signal a scarcity of massive Population III stars, suggesting the end of the period dominated by this stellar population.

## 4. Detectability of the He II and Hα Signals

We turn here to an analysis of the detectability of the Hα and He II signals in the intensity mapping regime, namely through measurements of the line-intensity power spectra. This is because sky-averaged measurements of the Hα and He II intensities of interest are challenging due to the prohibitively strong foregrounds and systematics. We begin by presenting the auto- and cross-power spectra of He II and Hα in Section 4.1. Next, we use these power spectra in Section 4.2 to estimate the signal-to-noise ratios $S/N_{H\alpha}$, $S/N_{He\,II}$, and $S/N_{H\alpha \times He\,II}$.

### 4.1. He II and Hα Power Spectra

The power spectra for Hα and He II were calculated from the fiducial simulation boxes described in Section 3. The left and middle columns of Figure 5 show the auto-power spectra for these two lines at $z = 7$ and $z = 13$, chosen for illustrative purposes. As expected, for the faintest IMFs, Hα increases in power from $z = 13$ to $z = 7$, reflecting the overall increase of the cosmic SFRD. The hardest IMFs, however, exhibit a small decrease in power, due to the larger Population III contribution at $z = 13$ compared to $z = 7$. This decrease is also observed for all six IMFs in the He II case, due to the strong sensitivity of this line to the Population III presence mentioned in the previous section. The He II sensitivity to Population III stars also manifests itself in the large differences in power for the different IMFs. For Hα, instead, the curves of the IMFs are almost indistinguishable in the top left panel because the Population III component contributes little to the power of that line at that redshift.

When taking into account the emission of the two lines separately as described above, the detectability of the line-intensity ratio will be limited by He II because this presents the weakest signal between the two lines. However, the line-intensity ratio detectability can be enhanced by considering the cross-correlation of both intensity fields. In brief, because both lines trace star formation, their signal will correlate, at least at linear (large) scales, and their cross-power will be higher than that of He II alone. In this case, the line-intensity ratio is obtained by dividing the cross-power by the power of Hα, where both observables now have a stronger signal than that of He II. Furthermore, considering the cross-power of the two lines aids in isolating the signal of interest from foreground or background contaminant interlopers (see, e.g., Beane et al. 2019). We come back to this point in Section 5. The rightmost column in Figure 5 shows the cross-power, where the signal levels are indeed higher than those of He II.

It is worth stressing again that we follow a simple approach for modeling the Population III star formation. The underlying assumption of Equation (7) is that Population III stars form in atomic cooling halos just like Population II stars, while some studies suggest that they may mainly reside in molecular cooling halos (Tan & McKee 2004; Bromm et al. 2009; Mebane et al. 2018). Because this difference may affect the power spectra, we explored this discrepancy and found that for a given fixed He II mean intensity, Equation (7) would result in a luminosity-averaged bias of Population III galaxies over-estimated by a factor of ~2. Since this is a small effect on the detailed shape of the power spectrum, compared to those introduced by different IMFs on the power spectrum amplitude, we do not perform more detailed analysis.

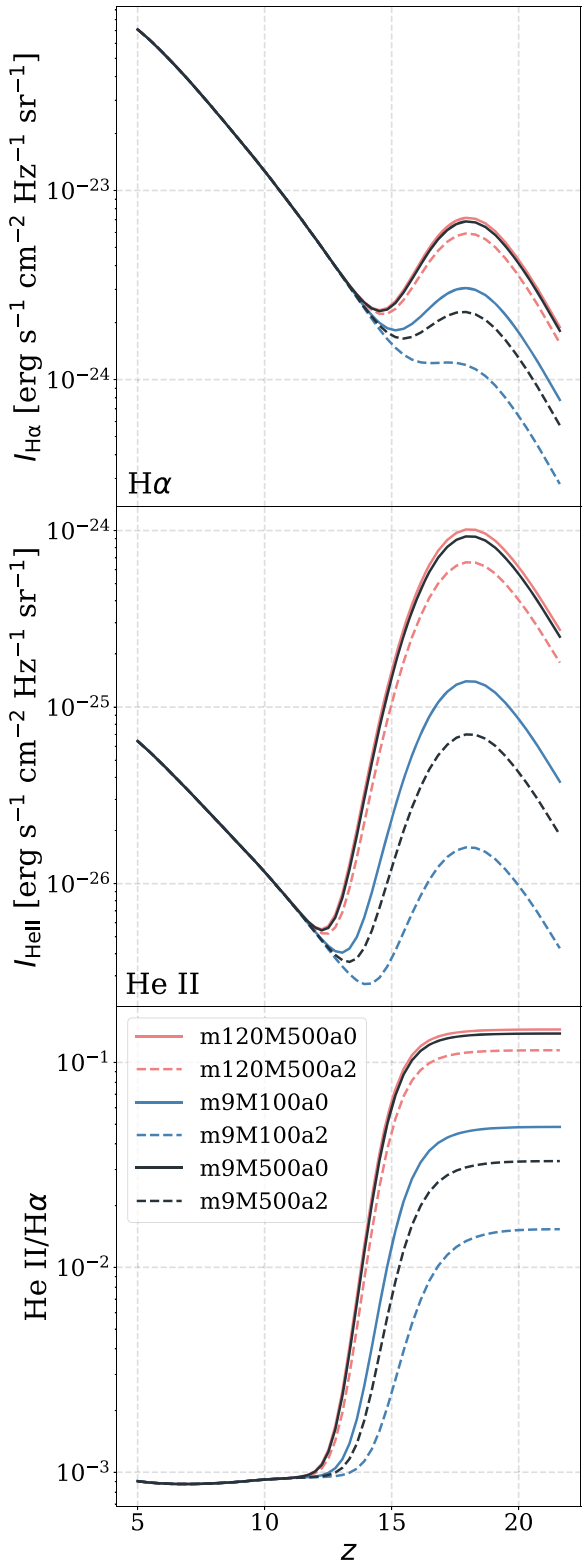

**Figure 4.** Mean line-intensity evolutions with redshift for Hα (top) and He II (middle), and their ratio (bottom), here using the alternative error function with center $z_t = 18$ and a width of $\sigma_p = 2$, for the six Population III IMFs described in Table 1. The shorter time span where the signal is dominated by Population III stars results in about one order of magnitude decrease in the maximum Population III intensity contribution, and highlights the need to go to high redshifts to perform Population III-dominated line-intensity ratio measurements.

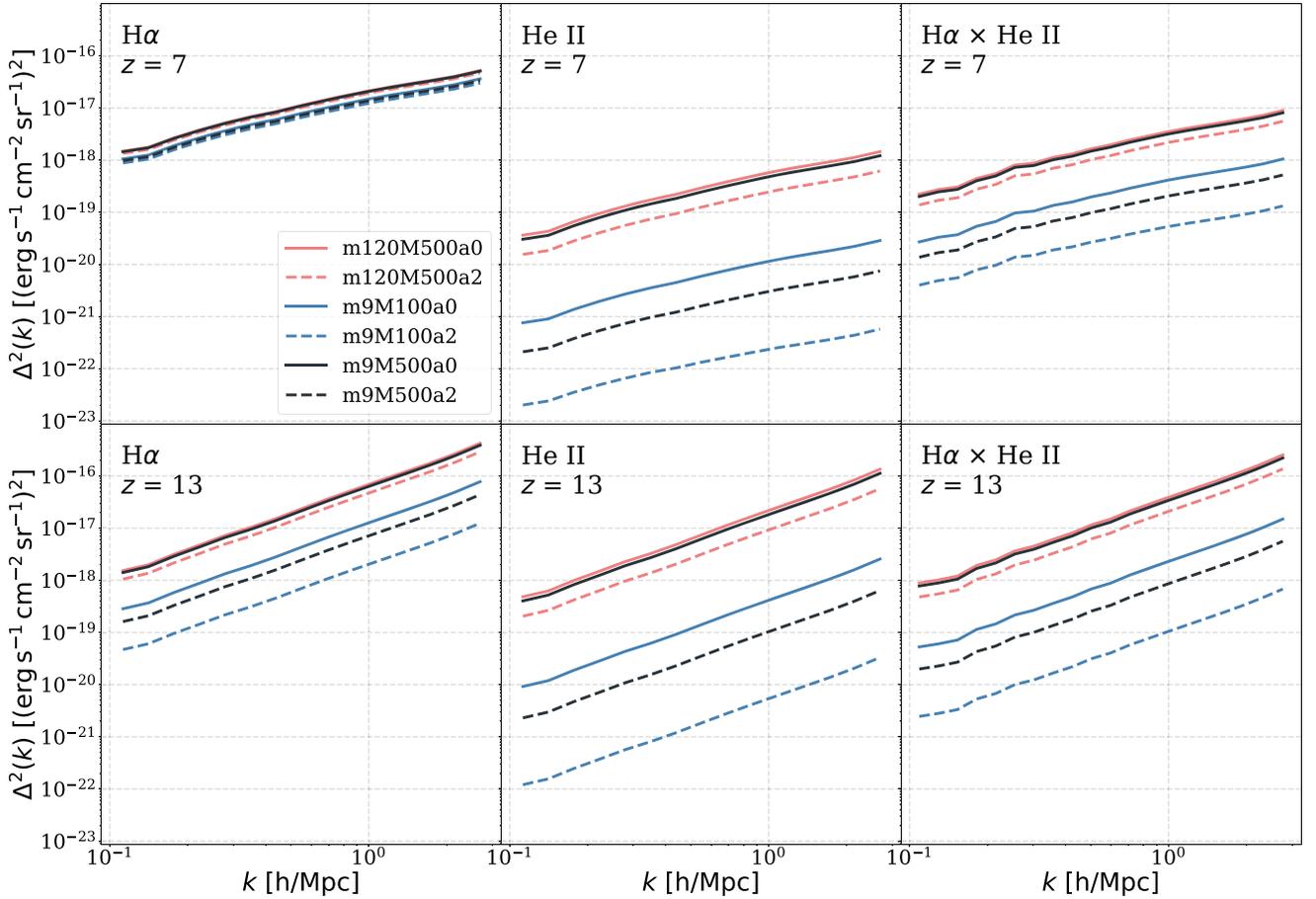

**Figure 5.** Auto-power spectra for Hα (left column), He II (middle column), and their cross-correlation (right column) at $z = 7$ (top row) and $z = 13$ (bottom row), with the six Population III IMFs described in Table 1. The hardest IMFs see a reduction of power at lower redshift due to the sensitivity of the signal to the presence of Population III stars. This effect is most notable for the case of the He II line.

### 4.2. S/N Estimation

With the power spectra for the fiducial Hα and He II calculations, we now estimate the signal-to-noise ratio (S/N) of these signals to determine their detectability. We explore two observational cases, designated as "present" and "improved," chosen to represent two different projections for next-generation experiments (see, e.g., Heneka & Cooray 2021). Both cases are based on proposed specifications of the Cosmic Dawn Intensity Mapper (CDIM) Probe Study (Cooray et al. 2019). Specifically, we assume a sky coverage of $\Omega_{tot} = 31$ deg$^2$ and a pixel size of 1″. Our "present" case has a surface brightness sensitivity of $\sigma_{noise} = 10^{-19}$ erg s$^{-1}$ cm$^{-2}$ sr$^{-1}$ Hz$^{-1}$ and a spectral resolving power of $R = 300$, while our "improved" case has sensitivity $\sigma_{noise} = 10^{-20}$ erg s$^{-1}$ cm$^{-2}$ sr$^{-1}$ Hz$^{-1}$ and resolution $R = 500$. With this set of parameters, the "present" case has observational capabilities similar to those of CDIM, while the "improved" case denotes an advanced instrument with about 10 times better sensitivity and about twice the spectral resolution.

The calculation of the S/N follows the method in Sun et al. (2019), to which we refer the reader for a more thorough description. For a given auto-power spectrum of interest, $P_{\nu\nu}(k)$, the uncertainty in its measurement can be expressed as

$$\delta P_{\nu\nu}(k) = \frac{P_{\nu\nu}(k) + P_{\nu\nu}^{noise}}{G(k)\sqrt{N_{modes}(k)}}, \quad (11)$$

where $G(k)$ denotes a smoothing factor due to finite spatial and spectral resolutions. The quantity $N_{modes}(k)$ refers to the number of independent Fourier modes, and $P_{\nu\nu}^{noise}$ denotes the power of thermal noise. $P_{\nu\nu}^{noise}$ can be expressed as

$$P_{\nu\nu}^{noise} = \sigma_{noise}^2 V_{vox}, \quad (12)$$

where $V_{vox}$ is the voxel size, which depends on the beam size $\Omega_{beam}$ and the spectral resolving power $R$. For a given cross-power spectrum, on the other hand, we use a modified version of Equation (11), such that

$$\delta P_{\nu\nu'}(k) = \frac{[P_{\nu\nu'}^2(k) + P_{\nu\nu}^{noise} P_{\nu'\nu'}^{noise}]^{1/2}}{G(k)\sqrt{2N_{modes}(k)}}, \quad (13)$$

where $\nu$ and $\nu'$ denote the frequency of the two lines. In the above equation, we neglect the error contribution from the auto-power spectrum. This is a valid approach because we have verified that the noise is several orders of magnitude larger than the signal, indicating that we are in the noise-dominated regime. Finally, the total S/N of the measured auto-power

spectrum is given by

$$\mathrm{S/N} = \sqrt{\sum_{k\,\mathrm{bins}} \left[\frac{P_{\nu\nu}(k)}{\delta P_{\nu\nu}(k)}\right]^2}, \quad (14)$$

replacing by the corresponding $\nu\nu'$ terms when computing the S/N for the cross-power. The $k$ range[9] of the summation runs from $k_{\min} \approx 0.1\,h/\mathrm{Mpc}$ to $k_{\max} \approx 3\,h/\mathrm{Mpc}$, as shown in Figure 5.

Figure 6 shows the S/N redshift evolutions of H$\alpha$ (top) and He II (middle) auto-power spectra and their cross-power spectra (bottom) for the "present" (thin lines) and "improved" (thick lines) cases, and for the six Population III IMFs described in Table 1. For both lines, the S/N is largest at redshifts where their respective emission signals are strongest in our fiducial model. This indicates that the S/N is mostly driven by the signal and not by the instrumental parameters. According to our fiducial model, the H$\alpha$ signal would be detectable for all IMFs and for both observational cases, except for the m9M100a2 IMF at the highest redshifts ($z \gtrsim 18$) in the "present" case. Indeed, even in the "present" case, $(\mathrm{S/N})_{\mathrm{H}\alpha} \gg 1$ for the majority of redshifts and IMFs. In the "improved" case, curves of the hardest IMFs asymptote to a plateau at S/N $\approx$ 4000–5000 rather than showing a clear peak, indicating the regime where sample variance dominates over the thermal noise and thus, where the S/N solely depends on $N_{\mathrm{modes}}$ instead of the signal level.

On the other hand, the power spectrum signal of He II is much weaker than that of H$\alpha$, resulting in a substantially lower S/N. For the He II signal in the "present" case, only the hardest IMFs (i.e., m120M500a0, m9M500a0, and m120M500a2) would be significantly detectable at $z \approx z_\mathrm{t} = 14$, where the He II emission peaks. However, for the "improved" case, these three hard IMFs would be detectable at all redshifts except $z \lesssim 7$, when the Population III SFRD becomes negligible, and the m9M100a0 and m9M500a2 IMFs would be detectable at $z \approx 14$. For this "improved" case, we note that these five IMFs of He II should always be detectable at the redshifts of interest for an He II/H$\alpha$ ratio measurement, since the peak in He II intensity corresponds to the redshift range dominated by Population III stars.

In the bottom panel of Figure 6, we observe the advantage of cross-correlating the weaker He II signal with the stronger H$\alpha$ signal. Cross-correlating boosts the observable signal at low redshifts, which, in the "improved" case, makes all the IMFs detectable even at $z \lesssim 7$. For the "present" case, the hardest IMFs (i.e., m120M500a0, m9M500a0, and m120M500a2) are significantly detectable at $z \gtrsim 7$, whereas for the He II signal considered alone, these IMFs were only detectable at $z \approx 14$. The m9M100a0 IMF is also detectable at $z \approx 14$ in this "present" case.

Finally, the S/N redshift evolution of our main observable, the He II/H$\alpha$ line-intensity ratio, is shown in Figure 7. As described in Section 4.1, the line-intensity ratio can be obtained by dividing the cross-power of He II and H$\alpha$ by the auto-power of H$\alpha$. Using this relation, we obtain the S/N values by propagating the errors on the power spectra shown in Figure 5. As with the S/N of the cross-correlation of He II and Hα shown in the bottom panel of Figure 6, all six IMFs in the "improved" case and the three hardest IMFs (i.e., m120M500a0, m9M500a0, and m120M500a2) in the "present" case are significantly detectable at virtually all redshifts. The m9M100a0 and m9M500a2 IMFs are detectable at $z \approx 14$ in the "present" case, and the m9M100a2 IMF remains undetectable.

---

[9] We note that the assumed survey specifications imply an accessible $k$ range slightly wider than the one used for our calculations at both $k_{\min}$ and $k_{\max}$ ends. However, we find a conservative estimate per the size and resolution of our simulation box is appropriate here, considering that lowest-$k$ modes tend to be overwhelmed by foreground and systematics, and that the line-intensity ratio of interest is most directly constrained by large-scale modes in the linear clustering regime.

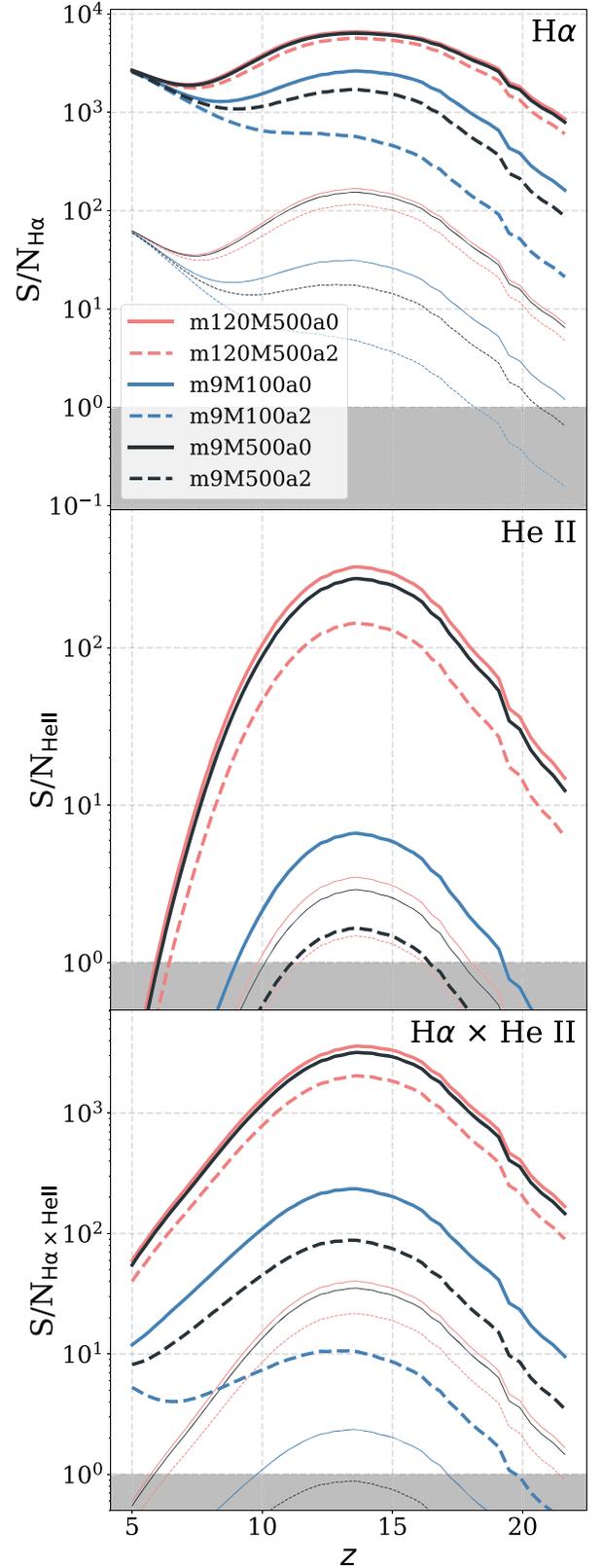

**Figure 6.** Signal-to-noise ratio (S/N) as a function of redshift for H$\alpha$ (top), He II (middle), and their cross-correlation (bottom), for the six Population III IMFs described in Table 1. The S/N values were calculated using the power spectra shown in 4.1, and two different hypothetical sets of instrument specifications. Our "present" case is shown as thin lines, while our "improved" case is shown as thick lines. The plots are grayed out below the S/N = 1 level.

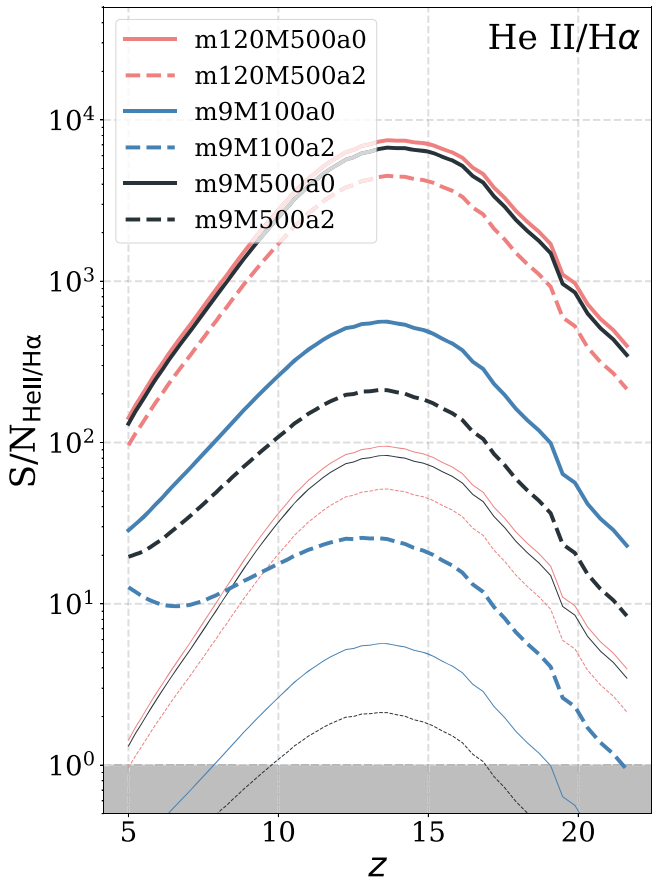

**Figure 7.** Signal-to-noise ratio (S/N) as a function of redshift of the He II/Hα line-intensity ratio for the six Population III IMFs described in Table 1. The S/N values were calculated using the power spectra shown in 4.1 and error propagation for two different hypothetical sets of instrument specifications. Our "present" case is shown as thin lines, while our "improved" case is shown as thick lines. The plots are grayed out below the S/N = 1 level.

In conclusion, we find that measurements of the He II/Hα ratio at a sufficiently high redshift to constrain the Population III IMF could be possible in the near future, assuming a dedicated deep survey and reasonable technological advances in instrumentation.

## 5. Discussion

We extend our discussion on the interpretation of the Hα and He II signals in Section 5.1 below, and highlight the challenges of line interloper separation in Section 5.2.

### 5.1. Interpretation of the Hα and He II Signals

As shown in Sections 3 and 4, the He II to Hα ratio is indeed a tracer for the Population III IMF, as suggested by Oh et al. (2001) and already visible in the early Population III work by Schaerer (2003). We have found that reasonable advances in instrument sensitivities would enable the measurement of the line-intensity ratio with sufficient S/N at redshifts $z \gtrsim 10$–12 when the signal is driven by massive Population III stars and top-heavy IMFs. In order to obtain constraints for less extreme Population III IMFs, however, one would ideally like to measure the ratio at multiple redshifts to recover the shape of the ratio evolution, as shown for example in the bottom panel of Figure 3. This is because recovering the shape of the ratio with redshift may reduce the degeneracy between the Population III IMF and the Population III SFRD. However, even knowing the shape of the ratio evolution, other sources of degeneracy may exist. Importantly, for a measurement of the ratio to be reliable, the contamination by Population II emission should be minimized, which requires information concerning the Population II stellar population at the corresponding redshift. Metal lines could prove useful in extracting this information, as mentioned by Visbal et al. (2015). The detection of oxygen lines such as [O II] 3727 Å or [O III] 5007 Å could point to the presence of Population II stars at a certain redshift, assuming that Population III stars are nearly metal free. In this case, one may try to remove the contribution from the Population II component or infer its SFRD to isolate the signal from Population III stars. Indeed, Moriwaki et al. (2018) demonstrated that optical metal lines such as [O III] 5007 Å can be useful probes of star-forming galaxies during reionization, and that narrow-band surveys of [O III] 5007 Å emitters by JWST NIRCam are expected to probe a wide luminosity range of the galaxy luminosity function. Furthermore, such surveys will provide better constraints on the ongoing formation of Population II stars than observations of FIR fine-structure lines (e.g., [C II] 158μm, [O III] 88μm) by ALMA, which are restricted to the very bright end of the luminosity function.

Meanwhile, it is worth noting that at $z \lesssim 10$, the He II signal could be contaminated by sources such as Population II Wolf–Rayet stars and high-redshift quasars, as already discussed by Visbal et al. (2015). If one were to measure the He II/Hα ratio at $z \lesssim 10$, then, these possible sources of contamination would have to be considered. For the case of quasars, spectral information concerning the width of the He II emission line may aid in uncovering their presence. This is because the line width is expected to be broader for quasars than for stars due to the high velocity dispersion of matter in the emission regions in quasars. In practice, however, it may be difficult to obtain such information from a large sky area covered by intensity mapping. We do not further examine these sources of contamination here because we primarily advocate for ratio measurements at $z \gtrsim 12$, but stress that initial estimates by Visbal et al. (2015) indicated that these contaminating sources may not be a major concern.

Additional information from the aforementioned observational probes and others such as, e.g., Galactic archeology, direct/indirect observations of high-redshift transient events such as long gamma-ray bursts, pair-instability supernovae, and direct-collapse black holes (Bromm 2013; Visbal et al. 2015; Mebane et al. 2018; Lazar & Bromm 2022) will be highly valuable to reduce the degeneracies mentioned here and to shed light on the IMF of Population III stars.

### 5.2. Line Interloper Separation

We note that the sensitivity analysis presented in Section 4 assumes a good control of instrumental and observational systematics, as well as foreground contamination. In particular, the residual contamination from low-redshift line interlopers after the separation of continuum foregrounds can be a challenge to the detection of high-redshift Hα and He II signals in autocorrelation. As discussed in Visbal et al. (2015), cross-correlating with another tracer, such as the H I 21 cm or CO line, at the same redshift as the target signal eliminates the line-confusion issue. However, measuring such cross-correlations at $z > 10$ can be very challenging given the intrinsic faintness of these alternative tracers. Meanwhile, as pointed out by Gong et al. (2014) and Heneka & Cooray (2021), only low-level mitigation via blind, high-flux masking is required to suppress the power of low-redshift line interlopers down to a reasonable level compared to that of the high-redshift signals, which tend to have significantly steeper line-luminosity functions thanks to the redshift evolution of the halo-mass function. This will be particularly true of high-redshift He II emission if it is primarily sourced by the highly abundant molecular cooling halos. Given the availability of imaging (and spectroscopic) data from deep, wide-field galaxy surveys to be conducted by, e.g., Euclid, the Vera Rubin Observatory, and the Roman Space Telescope soon, we expect the cleaning of interloping lines to be further facilitated by targeted masking based on the ancillary data from galaxy surveys (Sun et al. 2018).

For He II specifically, the presumably strong Lyα emission from Population III stars formed at a higher redshift can also be an important source of contamination, which is less likely to be removed by voxel masking due to the lack of ancillary data. In this case, de-confusion techniques based on the anisotropy of signal power spectra from redshift-space distortions can be leveraged to jointly fit multiple components from different redshifts (Gong et al. 2014, 2020; Cheng et al. 2016; Lidz & Taylor 2016).

Based on the various difficulties mentioned above, multi-line cross-correlations including our proposed approach of cross-correlating the He II and Hα signals may result in the simplest and most beneficial alternative to obtain clean, interloper-free measurements of the mean line intensity and ratio (Beane et al. 2019; Sun et al. 2019).

## 6. Summary And Conclusions

In this work, we have assessed the power of the line-intensity mapping technique to probe the initial mass function of Population III stars via the sky-averaged He II/Hα ratio of line emission. We have modeled the emission-line signals with a variant of the recently developed code LIMFAST (L. Mas-Ribas et al. 2022, in preparation; G. Sun et al. 2022, in preparation). Our code implements the galaxy-formation models developed by Furlanetto et al. (2017), and we considered a fraction of the star formation to be in the form of Population III stars through a simple redshift-dependent functional form. We computed the line emission using six Population III IMFs created following Mas-Ribas et al. (2016) and used as input spectra in the photoionization code Cloudy (Ferland et al. 2017). We also created and used a set of 41 metal-dependent Population II spectra of galaxies modeled with CloudyFSPS (Byler et al. 2017, 2018), as described in Section 2. In Section 3, we presented the evolution of the mean line emission and the line-intensity ratio with redshift, and we assessed the impact of model variations on the results. We then addressed the detectability of the signals in Section 4, through the computation of the auto- and cross-power spectra of the emission lines, and by considering two observational setups for an S/N estimate.

Mean line-intensity ratio measurements with values of He II/H $\alpha \gtrsim 0.1$ indicate the presence of Population III galaxies with stars of masses of several hundred solar masses ($\sim 500\,M_\odot$), and with top-heavy IMFs. We found, however, that this ratio value is reached when Population III stars dominate the signals, with no contamination from Population II stars, which, in our fiducial approach, implies redshifts of $z \gtrsim 10$–12. In this scenario, our detectability estimates indicate that a next-generation space mission with properties moderately more advanced than those of CDIM (Cooray et al. 2019) (10 times better sensitivity and about twice the spectral resolution) will be able to measure the He II power spectrum signal at S/N > 100 between $10 \lesssim z \lesssim 17$, and the Hα power spectrum at S/N better than $\approx 4000$, in a deep survey covering 31 deg². Furthermore, using the cross-correlation of the two lines provides better detectability levels than using the He II and Hα auto-power spectrum signals, especially when line interlopers are considered. These S/N values would indeed enable a reliable measurement of the aforementioned high line-intensity ratio value of He II/H $\alpha \gtrsim 0.1$. On the other hand, a measurement of He II/Hα $\lesssim 10^{-2}$ at any redshift would indicate a significantly small contribution of Population III galaxies to the total star formation, regardless of their IMF. This could be used to put limits on the duration of the impact of Population III stars to the overall star formation. However, a detection of the faint He II signal in these low-ratio cases may be only possible above the noise (S/N of a few) for top-heavy Population III IMFs, unless the He II signal is boosted by cross-correlating with the Hα signal. Finally, ratio values between He II/H $\alpha \sim 0.1$ and He II/Hα $\sim 10^{-2}$ are complex to interpret. They can be driven by various effects such as the Population III IMF (different upper mass bounds or slopes), by a significant fraction of Population II star formation reducing the true Population III ratio value, or by astrophysical effects such as the escape fraction of line photons or the galaxy duty cycle, among others. In this case, various measurements at different redshifts and the combination of the ratio information with other probes such as the detection of metal lines may be necessary to set tighter constraints on the Population III IMF. In summary, the He II/Hα ratio inferred from intensity mapping is a promising tracer for the elusive Population III IMF.

We thank the anonymous referee for comments that helped improve this paper. We also thank Steven Furlanetto, Adrian Liu, Adam Trapp, and Taj Dyson for useful discussions and comments on the manuscript. We acknowledge support from the JPL R&TD grant. This research was carried out at the Jet Propulsion Laboratory, California Institute of Technology, under a contract with the National Aeronautics and Space Administration.

*Software:* 21cmFAST (Mesinger & Furlanetto 2007; Mesinger et al. 2011), ARES (Mirocha et al. 2017), Cloudy (version 17.02, Ferland et al. 2017), LIMFAST (L. Mas-Ribas et al. 2022, in preparation; G. Sun et al. 2022, in preparation), CloudyFSPS (Byler et al. 2017, 2018).

## Appendix A
## Impact of Variations in the Ionization Parameter

For completeness, we have explored variations in the value of the ionization parameter used in our calculations given the sensitivity of the He II intensity to this quantity (e.g., Raiter et al. 2010). We have compared results from our fiducial ionization parameter combination of $\log U_{\text{Pop II}} = \log U_{\text{Pop III}} = -2$, to those with the combination $\log U_{\text{Pop II}} = -2$ and $\log U_{\text{Pop III}} = -1$. The small change in Population III ionization parameter does not result in significant differences for either H$\alpha$ or He II intensities (all changes were near or below the percent level). In another combination of ionization parameters, $\log U_{\text{Pop II}} = -4$ and $\log U_{\text{Pop III}} = -2$, H$\alpha$ was affected by a few percent given its mild sensitivity to changes in $U$. Notably, however, the line intensities of He II (and hence the ratio evolution) were visibly affected at low redshift where Population II stars dominate, as shown in Figure 8. For Population III IMFs with few massive stars, the line intensity of He II at low redshift, shown as thin lines, deviates downward from the fiducial curves, shown as thick lines, given the lower contribution from Population II stars. This deviation is also visible in the He II/H$\alpha$ ratio.

In sum, small variations in the ionization parameter for Population III stars appear to have little impact in the value of the line-intensity ratio. Variations in the ionization parameter for Population II stars, on the other hand, can affect the ratio evolution at low redshifts, where Population II stars dominate. This effect highlights the contaminant effect of Population II stars on the He II/H$\alpha$ ratio at low redshifts, and thus the need for ratio measurements to be at high redshift.

## Appendix B
## Ionization Histories

Despite the unknown nature of Population III stars, an important indirect constraint on their presence is the impact that they produce on the history of cosmic reionization. Figure 9 displays the ionization histories resulting from our computations using the six Population III IMFs, and the golden line denotes a Population II-only case with $f_{\text{PopIII}} = 0$ for comparison. The escape fraction of ionizing photons for Population II stars is set to $f_{\text{esc}}^{\text{II}} = 0.1$ in all cases, but we set the values for Population III stars to $f_{\text{esc}}^{\text{III}} = 0.1$ (top panel) and $f_{\text{esc}}^{\text{III}} = 0.01$ (bottom panel). A faster and earlier reionization is visible for most of the IMF cases with $f_{\text{esc}}^{\text{III}} = 0.1$ compared to the case of $f_{\text{esc}}^{\text{III}} = 0.01$ in the bottom panel. Furthermore, the inset plots show the values of the total electron optical depth $\tau_e$ for each IMF, with gray regions indicating values beyond $2\sigma$ from the Planck estimate $\tau_e = 0.0627^{+0.0050}_{-0.0058}$ (de Belsunce et al. 2021). When Population III stars are included in the calculations and an escape fraction $f_{\text{esc}}^{\text{III}} = 0.1$ is assumed, only the m9M100a2 Population III IMF produces results consistent with the Planck value. An escape fraction of $f_{\text{esc}}^{\text{III}} \lesssim$ of a few percent is required for the other Population III IMFs, as modeled in the fiducial case, to be consistent with Planck results. The exact escape fraction values allowed by these constraints depend on the specific IMF, since the more top-heavy the IMF, the smaller the permitted escape fraction as result of the higher stellar ionizing efficiency. However, we stress that the reionization picture discussed here also depends strongly on the Population III SFRD, which remains almost unconstrained observationally, and which we have modeled with a simple approach (Section 2.1.1). Meanwhile, because massive Population III stars can grow larger Strömgren spheres and provide stronger radiative and supernova feedback to carve out ionized channels to the IGM, larger values of $f_{\text{esc}}^{\text{III}}$ might also be plausible (Wise et al. 2014; Tanaka & Hasegawa 2021). These larger values, however, are beyond the scope of this paper. In summary, the above calculations suggest that the escape fraction of our top-heavy Population III IMFs should be small to be consistent with recent reionization history constraints. Most significantly, the differences between the two escape fraction values considered have almost negligible effect on the line-emission and line-intensity ratios of interest here.

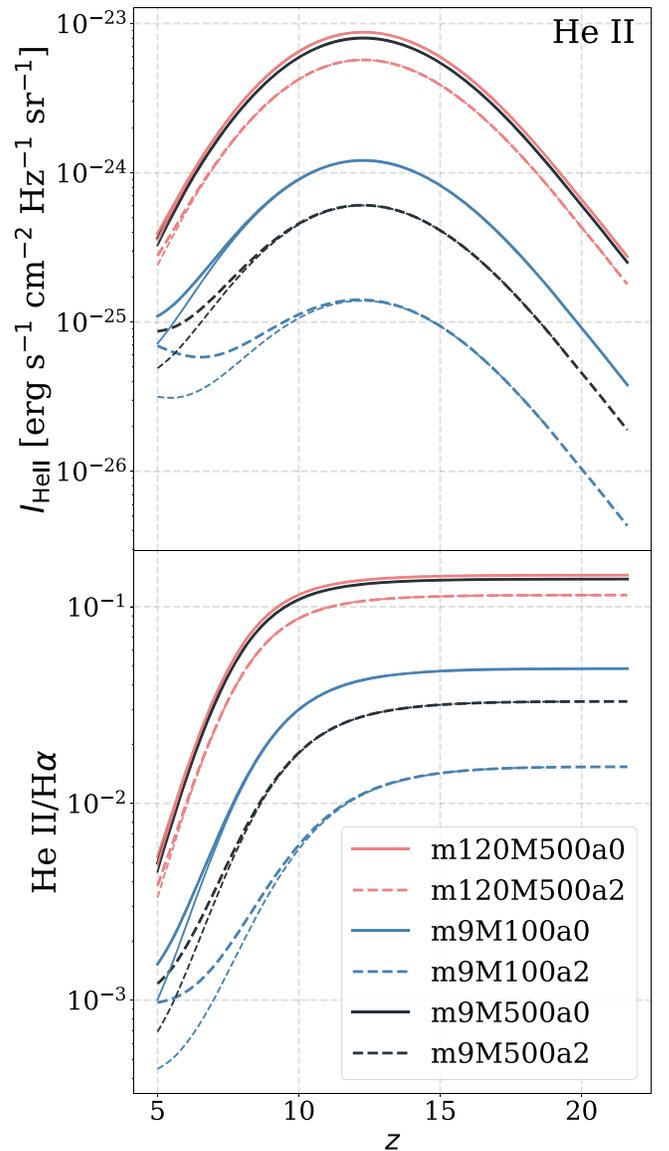

**Figure 8.** Simulated line intensities of He II (top) and the He II/H$\alpha$ ratio (bottom) with an ionization parameter combination of $\log U_{\text{Pop II}} = -4$ and $\log U_{\text{Pop III}} = -2$, plotted as thin curves, for the six Population III IMFs described in Table 1. The fiducial counterparts to these curves (with an ionization parameter combination of $\log U_{\text{Pop II}} = -2$ and $\log U_{\text{Pop III}} = -2$) are also plotted in thicker lines. This change in $\log U_{\text{Pop II}}$ only causes deviations at low redshifts, since the decline in the Population III population is not as immediately counteracted by the rise in the Population II population.

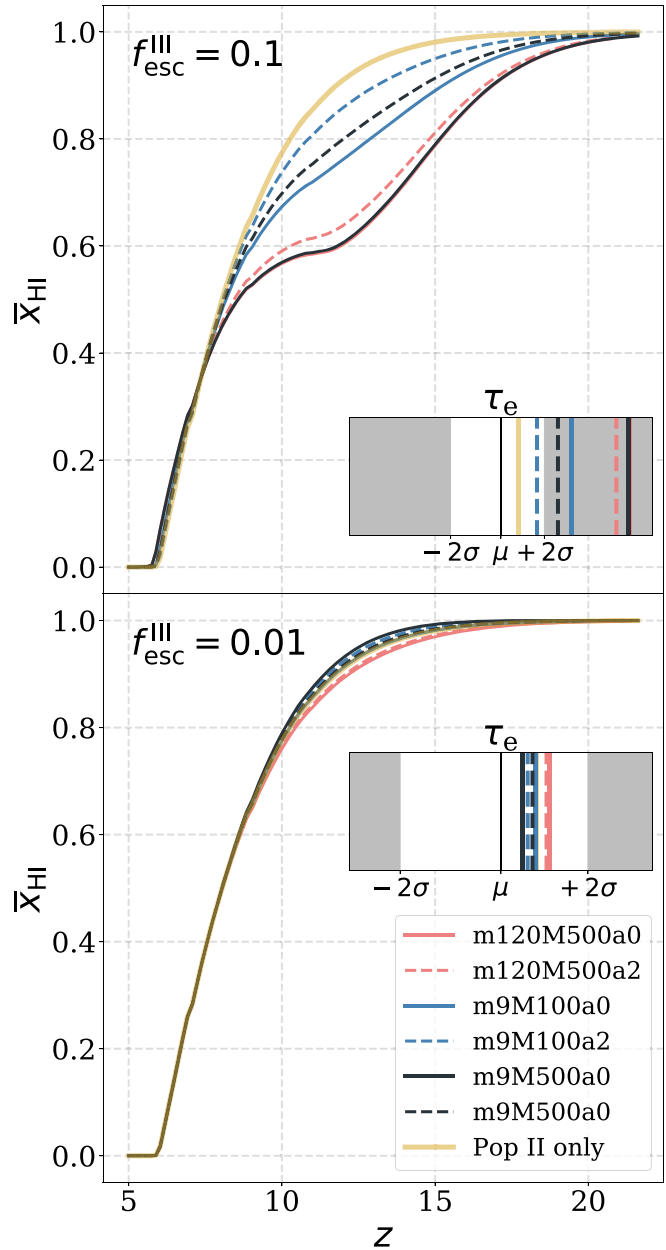

**Figure 9.** Ionization histories with Population II escape fraction of $f_{esc}^{II} = 0.1$, and Population III escape fractions $f_{esc}^{III} = 0.1$ (top) and $f_{esc}^{III} = 0.01$ (bottom), for the six Population III IMFs described in Table 1. The thick golden curve corresponds to a Population II-only scenario, i.e., with $f_{PopIII} = 0$. The inset plots show the values of the total electron optical depth $\tau_e$ for each IMF, with gray regions indicating values beyond $2\sigma$ from the Planck estimate $\tau_e = 0.0627^{+0.0050}_{-0.0058}$ (de Belsunce et al. 2021). When Population III stars are included in the calculations and an escape fraction $f_{esc}^{III} = 0.1$ is assumed, only the m9M100a2 Population III IMF produces results consistent with the Planck value. An escape fraction of $f_{esc}^{III} \lesssim$ a few percent is required for the other Population III IMFs, as modeled in the fiducial case, to be consistent with the Planck results, the exact number depending on the specific IMF.

**ORCID iDs**

Jasmine Parsons ⓘ https://orcid.org/0000-0002-6013-4655
Lluís Mas-Ribas ⓘ https://orcid.org/0000-0003-4584-8841
Guochao Sun ⓘ https://orcid.org/0000-0003-4070-497X